\documentclass[letterpaper]{article}
\usepackage{arxiv} 
\usepackage{times} 
\usepackage{helvet} 
\usepackage{courier} 
\usepackage[hyphens]{url} 
\usepackage{graphicx}  
\usepackage{caption}  
\usepackage{algorithm}
\usepackage{algorithmic}
\usepackage{newfloat}
\usepackage{listings}
\usepackage[utf8]{inputenc}
\usepackage{booktabs}
\usepackage{tabularx}
\usepackage{pifont}
\usepackage{placeins}
\usepackage[font=small,labelfont=bf]{caption}
\usepackage{csquotes}
\usepackage{subfiles}
\usepackage{afterpage}
\usepackage{authblk}
\usepackage{hyperref}

\usepackage[backend=biber,style=numeric,sorting=none]{biblatex}
\DeclareCaptionStyle{ruled}{labelfont=normalfont,labelsep=colon,strut=off}
\floatstyle{ruled}
\newfloat{listing}{tb}{lst}{}
\floatname{listing}{Listing}

\renewcommand{\headeright}{}
\renewcommand{\shorttitle}{}

\author[1, *]{Patricia~Paskov}
\author[1]{Kevin~Wei}
\author[2]{Shen~Zhou~Hong}
\author[3]{Dan~Bateyko}
\author[4]{Xavier~Roberts-Gaal}
\author[1]{Carson~Ezell}
\author[5]{Gailius~Praninskas}
\author[6]{Valerie~Chen}
\author[7]{Umang~Bhatt}
\author[1]{Ella~Guest}

\affil[1]{\href{https://rand.org/}{RAND}, Santa Monica, CA 90407, United States}
\affil[2]{\href{https://jhu.edu/}{Johns Hopkins University}, Baltimore, MD 21218, United States}
\affil[3]{\href{https://cornell.edu/}{Cornell University}, Ithaca, NY 14853, United States}
\affil[4]{\href{https://harvard.edu/}{Harvard University}, Cambridge, MA 02138, United States}
\affil[5]{\href{https://lse.ac.uk/}{London School of Economics}, London, WC2A 2AE, United Kingdom}
\affil[6]{\href{https://cmu.edu/}{Carnegie Mellon University}, Pittsburgh, PA 15213, United States}
\affil[7]{\href{https://cam.ac.uk/}{University of Cambridge}, Cambridge, CB2 1TN, United Kingdom}
\affil[*]{Address correspondence to \href{mailto:ppaskov@rand.org}{\texttt{ppaskov@rand.org}}}

\title{RCTs for Frontier AI Governance: Methodological Challenges and Solutions for Human Uplift Studies}

\begin{document}

\maketitle

\begin{abstract}
Human uplift studies, or studies that measure the effects of AI access on human performance via randomized controlled trials (RCT) or similar methodologies, increasingly inform frontier AI governance and deployment decisions. While RCT methods are robust in other fields, their interaction with the distinctive properties of frontier AI systems remains underexamined, particularly when results are used to inform high-stakes decisions. We present findings from interviews with 16 expert practitioners with experience conducting human uplift studies in domains including biosecurity, cybersecurity, education, and labor. Across interviews, experts described a recurring tension between the standard causal inference assumptions upon which human uplift studies rely and the object of study itself. Rapidly evolving AI systems, shifting baselines, heterogeneous and changing user proficiency, and porous real-world settings strain assumptions underlying internal, external, and construct validity, complicating the interpretation and appropriate use of uplift evidence. We contribute (1) a synthesis of methodological challenges in human uplift studies, mapped to risks to study validity and classified by their degree of specificity to large language model (LLM) systems, and (2) a mapping from challenges to proposed solutions. By collating expert-identified challenges and solutions, we seek to clarify the interpretive limits and appropriate uses of human uplift evidence, to align evaluation practice with the decisions it informs, and to support more coordinated methodological foundations for AI governance.
\end{abstract}

\newpage
\setcounter{tocdepth}{3}
\setcounter{secnumdepth}{3}
\tableofcontents

\twocolumn

\section{Introduction}
\begin{table*}[!t]
\centering
\footnotesize
\setlength{\tabcolsep}{3pt}
\renewcommand{\arraystretch}{1.15}
\newcommand{\mapx}{\ding{51}}
\begin{tabularx}{\textwidth}{p{4.4cm} *{9}{>{\centering\arraybackslash}X}}
\toprule
&
\multicolumn{9}{c}{\textbf{Methodological Challenges (by Research Phase)}} \\
\cmidrule{2-3} \cmidrule{4-5} \cmidrule{6-8} \cmidrule{9-10}
& \multicolumn{2}{c}{\textbf{Design}} 
& \multicolumn{2}{c}{\textbf{Recruitment}} 
& \multicolumn{3}{c}{\textbf{Execution}} 
& \multicolumn{2}{c}{\textbf{Documentation}} \\
\cmidrule(lr){2-3} \cmidrule(lr){4-5} \cmidrule(lr){6-8} \cmidrule(lr){9-10}
\textbf{Solutions} &
\rotatebox{45}{C1 Measurement [C] $\star$} &
\rotatebox{45}{C2 Controls [C] $\star\star$} &
\rotatebox{45}{C3 Recruitment [E] $\star$} &
\rotatebox{45}{C4 AI Literacy [I, E] $\star\star$} &
\rotatebox{45}{C5 Intervention Fidelity [C, I] $\star\star\star$} &
\rotatebox{45}{C6 Interference [I] $\star\star$} &
\rotatebox{45}{C7 Expectancy Effects [I, E] $\star\star$} &
\rotatebox{45}{C8 Documentation [E] $\star$} &
\rotatebox{45}{C9 Interpretation [E] $\star$} \\
\midrule
S1 Standardized Task Libraries 
& \mapx &  &  &  &  &  &  & \mapx & \mapx \\
S2 Baseline \& Control Conventions 
&  & \mapx &  &  &  & \mapx &  & \mapx & \mapx \\
S3 Leveling AI Literacy 
&  &  &  & \mapx &  &  &  & \mapx & \mapx \\
S4 Versioned Snapshots 
&  &  &  &  & \mapx &  &  & \mapx & \mapx \\
S5 Interference Management 
&  &  &  &  &  & \mapx &  & \mapx & \mapx \\
S6 Natural Experiments \& Phased Rollouts 
&  &  & \mapx &  &  &  &  &  & \mapx \\
S7 AI-Accelerated Research Methods 
&  &  & \mapx & \mapx & \mapx & \mapx &  &   & \mapx \\
S8 Post-Hoc Analysis 
&  &  & \mapx & \mapx & \mapx & \mapx & \mapx &  & \mapx \\
S9 Information Security Frameworks 
&  &  &  &  &  &  &  & \mapx & \mapx \\
\bottomrule
\end{tabularx}
\vspace{1mm}
\caption{Methodological challenges and solutions in human uplift studies: a mapping between proposed solutions (rows) and methodological challenges (columns) identified in 16 expert interviews. Challenges are grouped by the primary research phase in which they arise, tagged by the primary form of validity they threaten (construct [C], internal [I], external [E]), and rated for LLM-specificity (from familiar RCT threats intensified by LLMs ($\star$), to those with RCT precedent but inadequate standard mitigations ($\star\star$), to challenges native to LLM-based systems ($\star\star\star$)). Mapping is illustrative rather than exhaustive.}\label{tab:challenge_solution_map}
\end{table*}

Large language model (LLM) systems are reshaping fundamental aspects of human society, from how we work and learn to how we make decisions about national security and public policy \cite{bengio_international_2026,maslej2025artificialintelligenceindexreport}. Society's ability to anticipate, manage, and govern these transformations depends in part on how well we can evaluate AI's real-world impacts. Governance frameworks increasingly rely on evaluations as a precondition for responsible deployment: the EU AI Act mandates evaluations of frontier models \cite{euaiact}; US state-level laws such as New York's RAISE Act and California's SB 53 require public reporting of safety evaluations \cite{raise_act_2025, sb53_2025}; and frontier developers voluntarily commit to evaluation-based risk thresholds through Responsible Scaling Policies and Frontier Safety Frameworks \parencite[e.g.,][]{anthropic_anthropics_2023, deepmind_fsf_2024, openai_preparedness_2025}. Yet the evaluation methods on which this paradigm relies largely compare AI systems with each other, and often fall short in measuring how systems impact users and society in practice \cite{schwartz2025realitychecknewevaluation, ibrahim_towards_2025, weidinger2023sociotechnicalsafetyevaluationgenerative, Strauss_2025}.

Human uplift studies offer a way to close this gap. By measuring the extent to which access to or use of LLM systems shifts human performance on a task relative to a counterfactual through randomized controlled trial (RCT) or similar methodologies, uplift studies provide rigorous causal evidence about AI's effects on the people who interact with it \cite{frontier_model_forum_frontier_2025,uk_ai_security_institute_ai_2024}. Human uplift studies are increasingly discussed by international consortia \cite{bengio_international_2024, frontier_model_forum_frontier_2025}, governments \cite{ec_code_2025, uk_ai_security_institute_early_2024, uk_ai_security_institute_ai_2024, us_ai_safety_institute_nist_2024}, developers \cite{anthropic_system_2025, patwardhan_building_2024, shah_approach_2025, shanghai_ai_lab_frontier_2025}, and AI evaluation contexts \cite{mccaslin_stream_2025, paskov_preliminary_2025, metr_what_2025, future_of_life_institute_2025_2025, janjeva_evaluating_2024, bucknall_which_2025, wilson_us_2024, karnofsky_if-then_2024}.\footnote{The term \enquote{human uplift study} has emerged as a consensus label within AI policy and evaluation discourse to describe this class of methods \cite{bengio_international_2024, frontier_model_forum_frontier_2025, uk_ai_security_institute_ai_2024}. For clarity, we use \enquote{human uplift study} to refer broadly to RCT-based or quasi-experimental evaluations of AI’s effects on human task performance, while acknowledging that individual studies may vary in design details and rigor. We note that the term itself is a misnomer: not all such studies find that access to frontier AI systems increases human performance, e.g., \textcite{becker_measuring_2025}.}

This promise, however, outpaces methodological understanding. While RCT methodology is well-established in fields including medicine, economics, and human-computer interaction  \cite{eble_minimizing_nodate, tong_just_2023, dahlinger_impact_2018, jonesPeopleCannotDistinguish2025}, its application to frontier AI systems poses new challenges. Moreover, study results are not always made public; even when published, results often omit key methodological details due to safety concerns, such as in \textcite{anthropic_system_2025}. As uplift studies increasingly shape consequential decisions about AI deployment, safety, and regulation, gaps between the methodological foundations of this evidence and the institutional conditions under which it is produced and used become questions not only of research design but of AI governance.

This article fills that gap with findings from 16 interviews of experts who have conducted or were contemporaneously conducting human uplift studies involving large language models (LLMs). Our main contributions are: (1) a synthesis of methodological challenges in LLM uplift studies, mapped to threats to construct, internal, and external validity and classified by their degree of LLM-specificity; and (2) a mapping from challenges to expert-proposed solutions. Our focus and analysis are centered on human uplift studies involving LLM-based systems, though we occasionally use the broader term ``AI systems'' for generality. By collating expert-identified challenges and solutions, we seek to clarify the interpretive limits and appropriate uses of human uplift evidence, and to promote the development of increasingly rigorous and robust AI evaluations for governance.

\section{Background \& related work}

Evaluations provide stakeholders with critical insights into the capabilities, risks, and opportunities of frontier AI systems \cite{bengio_international_2026,burden_paradigms_2025, anthropic_system_2025,noauthor_responsible_nodate, uk_ai_security_institute_ai_2024, ec_code_2025}. Current approaches, including multiple-choice question-answer (MCQA) benchmarks, red-teaming, and long-form agent evaluations, present distinct trade-offs in validity, reproducibility, and resource requirements \cite{mouton_operational_2024,uk_ai_security_institute_ai_2024,frontier_model_forum_frontier_2025, reuel2024betterbenchassessingaibenchmarks, lin2024achillesheelsurveyred, burden_paradigms_2025, wei2025methodological}. While MCQA benchmarks provide structured performance measurement, they often neglect system interaction with users or environments \cite{weidinger2023sociotechnicalsafetyevaluationgenerative, schwartz2025realitychecknewevaluation, ibrahim_towards_2025, Strauss_2025}. As such, benchmarks alone poorly predict downstream impacts, especially on economically or strategically important tasks~\cite{schaeffer_why_2025, brooks_is_2025, narayanan_gpt-4_2025, ho_real_2025}. Red-teaming approaches, while involving human-computer interaction, often lack the controlled structure for reliable causal estimates \cite{feffer_red-teaming_2025, friedler_ai_2023}.

Human uplift studies use RCT or similar methodology to measure the causal impacts of AI systems on human performance, drawing upon decades of experimental rigor from medicine, economics, and social science \cite{ibrahim_towards_2025,dobbe2021hardchoicesartificialintelligence,weidinger2023sociotechnicalsafetyevaluationgenerative,us_ai_safety_institute_nist_2024,Farrell2025-th, eble_minimizing_nodate, frontier_model_forum_frontier_2025, hoBioriskEvaluationsAI2025a}. RCTs originated in medical research as the gold standard for establishing causal relationships between interventions and outcomes, with rigorous standards developed over decades to minimize bias and ensure reliable inference \cite{eble_minimizing_nodate}. RCT methodologies have been adopted across fields including human-computer interaction \cite{tong_just_2023, dahlinger_impact_2018, price_engaging_2020, jeong_huggable_2018, milliCausalInferenceStruggles2022, lee_learning_2025, jakesch_ai-mediated_2019, g_mitchell_reflection_2021, bassen_reinforcement_2020} and economics and social sciences \cite{banerjee_field_2020, banerjeeInfluenceRandomizedControlled2020, angrist_mostly_2009}. Across these domains, RCTs are grounded in the \textit{Potential Outcomes Framework} , which specifies the assumptions required for causal inference and unbiased treatment-effect estimation \cite{rubin1974estimating, imbens2015causal}. 

In recent years, researchers have leveraged RCTs and other methods to evaluate the impact of LLMs on human performance across a growing set of domains, including biological threats \cite{mouton_operational_2024, patwardhan_building_2024,krishnaEvaluatingCriticalRisks2025, xai_grok_2025, zhangLLMNoviceUplift2026}, developer productivity \cite{becker_measuring_2025, chen_code_2025}, legal services \cite{schwartz2025realitychecknewevaluation, choiAIAssistanceLegal2024}, customer service \cite{brynjolfsson2025generative}, and academic research \cite{brodeur_comparing_2025, ratkovicHarnessingGPTEnhanced2025}. 
This breadth spans both benefit-oriented questions, such as productivity gains, and worst-case misuse risks. As decision-makers and policymakers increasingly turn to human uplift studies to inform deployment and governance decisions, the methodological soundness of these studies carries more weight. 

The application of RCTs to LLMs calls into question the core assumptions of the \textit{Potential Outcomes Framework} \cite{rubin1974estimating, imbens2015causal}, including stable interventions and well-defined counterfactuals. While violations of these assumptions are documented even in mature fields \cite{eble_minimizing_nodate}, LLM-based systems place additional and distinctive pressures on these assumptions. Rapid model iteration can undermine intervention fidelity over the course of a study, while the widespread integration of AI tools into everyday workflows \cite{menlo_2025_state_genai, mckinsey_2025_state_ai} complicates the specification of controls and the prevention of contamination.\footnote{Intervention fidelity refers to whether the treatment actually delivered matches the treatment specified in the study design. In the AI context, this is threatened when models update, safety filters change, or system configurations shift during a study. For more discussion, see C5.} 

Amid these methodological questions and an evolving empirical landscape, much of the practical experience of conducting human uplift studies remains under-documented. Many studies are conducted under security constraints or commercial confidentiality \cite{10.1145/3375627.3375815, Strauss_2025, delaney2024mappingtechnicalsafetyresearch, mccaslin_stream_2025}, limiting transparency about design trade-offs, failure modes, and interpretive pitfalls. As a result, stakeholders often rely on human uplift evidence without a clear view of where standard assumptions hold, where they break down, and how researchers adapt in practice. In this article, we seek to fill this transparency gap. 

\section{Methodology}

We employ expert interviews as the primary method for this study. Human uplift studies, when published, often only report high-level results and frequently omit details on methodological trade-offs, execution challenges, and interpretive uncertainties. Expert interviews allow us to surface this tacit  knowledge directly, including insights from studies that remain unpublished or delayed due to security or commercial constraints. This project was reviewed and deemed exempt by our organization’s Institutional Review Board.

\subsection{Interview protocol} \label{subsec:Methods:Interviews}

We conducted semi-structured expert interviews between July and August 2025 with experts who had conducted or were, at the time of interview, conducting  at least one human uplift study of an LLM-based system. We selected expert participants using a snowball sampling method \cite{parker_snowball_2019}, seeded with a rapid literature review of published human uplift studies involving LLMs that fulfilled the criteria in Table~\ref{tab:Inclusion_Exclusion_Criteria}. For full sampling procedures, refer to Appendix~\ref{appendix:Systematic_Review}. In total, we reached out to 53 experts and secured participation of 16 interviewees (32.08\% participation rate). Interviews were scheduled for 60 minutes and lasted between 30--70 minutes. The semi-structured script addressed demographics, human uplift study history, methodological challenges and solutions, and open-ended questions. The methodology and script are further detailed in Appendix~\ref{appendix:Interview_Methodology} and~\ref{appendix:Script}, respectively. 

\subsection{Thematic analysis} \label{subsec:Methods:Analysis}

Following interview completion and transcription, we conducted a qualitative thematic analysis \cite{braun_using_2006} to identify high-level methodological challenges and solutions. We used a two-stage inductive approach, treating coding as an interpretive and reflexive process \cite{braun_toward_2023} to identify themes from the bottom up. The final codebook contained 30 codes across 7 categories (Appendix~\ref{appendix:Codes}), which we organized into broader themes (the Results section). Through thematic analysis, we aimed to identify high-level challenges and potential solutions surfaced in expert interviews. Each transcript was independently coded by two annotators. Following established qualitative research practices \cite{mcdonald_reliability_2019, braun_thematic_2024, liUnderstandingChallengesDevelopers2022, wei_how_2024}, we did not calculate inter-rater reliability, given our goal to interpret core narratives rather than make objective or predictive claims.

\subsection{Validity mapping}
\label{subsec:Methods:ValidityMapping}

We map challenges to three established forms of validity: construct, internal, and external validity (Table \ref{tab:validity_definitions}). This typology derives from \cite{campbell_1957} and subsequent work \cite{campbell1979quasi, cook2002experimental} and complements recent work in AI evaluation adopting and extending validity frameworks to assess the reliability and generalisability of model evaluations \cite{salaudeen_measurement_2025, chouldechova2024sharedstandardvalidmeasurement}. \footnote{Beyond construct, internal, and external validity, the Campbell tradition includes a fourth category, statistical conclusion validity, which we omit because practitioner-reported challenges concerned the definition, identification, and generalization of causal effects rather than statistical estimation. The Campbell framework has been widely adopted across the social sciences: political science adopted all four validity types \cite{morton2010experimental}, while economics' ``credibility revolution'' \cite{meyer1995natural, angrist2010credibility} draws on the same tradition, emphasizing internal and external validity while largely dropping construct validity from its methodological vocabulary. While there is no single universally-accepted validity classification, we adopt the Campbell typology as a minimal established framework that cleanly maps to the challenges emerging from interviews. We retain construct validity as a distinct category because many challenges practitioners face in AI evaluation concern whether their measures capture what they intend to measure. We consider this a question logically prior to, and distinct from, whether internal causal identification assumptions hold.} Mapping was independently completed by Author ~1 and Author ~2, with a third author adjudicating disagreements to reach consensus. Used as an analytic lens, this framework helps to identify when and how inferential claims are at stake; and seeks to foster more well-informed design and interpretation of human uplift studies.  

\begin{table*}[t]
\centering
\small
\begin{tabular}{p{0.12\linewidth} p{0.82\linewidth}}
\toprule
\textbf{Validity type} & \textbf{Definition and common threats} \\
\midrule
\textbf{Construct [C]} &
The extent to which study operations (e.g., interventions, measures, settings, participants) correspond to intended abstract constructs \cite{campbell1979quasi, cook2002experimental, salaudeen_measurement_2025}. Threats arise when the treatment is ill-defined (e.g., the LLM intervention updates mid-study), when control conditions fail to represent the relevant counterfactual, or when outcome measures capture only a subset of pathways relevant to the decision-relevant construct. \\[0.5em]

\textbf{Internal [I]} &
The extent to which the assumptions required to identify a causal effect between the explanatory variable and the outcome of interest are satisfied within the study context \cite{imbens2015causal, cook2002experimental}. Threats arise when treatment diffuses across experimental boundaries through spillovers or contamination, or when participants are differentially exposed to varying versions of an intervention. \\[0.5em]

\textbf{External [E]} &
The extent to which internally valid causal effects generalize to different individuals, contexts, and outcomes \cite{meyer_natural_1995, cook2002experimental}. Threats arise when the recruited sample diverges from the decision-relevant population, when user proficiency co-evolves with the technology, or when baselines shift over time. \\
\bottomrule
\end{tabular}
\caption{Validity dimensions used to organize methodological challenges in human uplift studies.}
\label{tab:validity_definitions}
\end{table*}

\subsection{LLM-specificity mapping}
We further map each challenge along a three-level spectrum reflecting the degree to which it emerges due to the distinctive properties of LLM-based systems (Table~\ref{tab:llm_specificity}). Mapping was independently completed by Author~2 and Author~6, with a third author adjudicating disagreements to reach consensus. 

\begin{table*}[h]
\centering
\small
\begin{tabular}{p{0.1\linewidth} p{0.82\linewidth}}
\toprule
\textbf{Level} & \textbf{Definition} \\
\midrule
\textbf{LLM-native}
$\star\star\star$ 
& The causal mechanism generating the challenge is specific to properties of LLM-based systems; precedents in prior RCT contexts are weak or absent. \\
\addlinespace
\textbf{LLM-transformed}
$\star\star$ 
& Precedent exists in prior RCT contexts, but standard mitigations are inadequate for preserving the relevant form(s) of validity in LLM studies. \\
\addlinespace
\textbf{LLM-intensified}
$\star$ 
& Precedent exists in prior RCT contexts, and standard mitigations remain largely adequate; LLMs primarily increase the frequency or severity of the threat. \\
\bottomrule
\end{tabular}
\caption{Three-level classification of how distinctly each methodological challenge emerges from properties of LLM-based systems.}
\label{tab:llm_specificity}
\end{table*}

\subsection{Limitations} \label{subsec:Methods:Limitations}

Our interview process was constrained by sample size and by potential sampling and response bias. Given the nascency of LLM human uplift research and the small population of experts in this area, however, our sample of 16 experts is in line with norms in expert interview research in AI ethics \cite{wei_how_2024, wangStrategiesIncreasingCorporate2025, rostamzadeh_healthsheet_2022, mehandruReliableSafeUse2022, cheongAIAmNot2024, 10.1145/3630106.3658926, 10.1145/3630106.3658959, 10.1145/3531146.3533113, 10.1145/3630106.3658988} and human-computer interaction \cite{agrawalExploringDesignGovernance2021, zhangIEDSExploringIntelliEmbodied2025, liUnderstandingChallengesDevelopers2022, sekwenzItUnfairIt2025, burukDesigningPlayfulBodily2023}.\footnote{The population of experts with direct experience conducting LLM human uplift studies is small; in later rounds of our snowball sampling, interviewees frequently identified experts already on our outreach list, suggesting we were approaching the boundary of the reachable expert population. As one expert noted, ``the AI uplift world, it's still kind of small.'' Yet content saturation in similar expert studies has been achieved with a dozen or fewer interviews \cite{hennink_sample_2022, guest_how_2006}. We observed that later interviews surfaced few new themes or challenges beyond those identified in earlier interviews, consistent with content saturation, though we did not conduct a formal saturation analysis. Our 32\% participation rate was comparable to that of similar studies \cite{harrap_randomised_2023}.} Our interview sample reflects broader biases representative of the field: experts were predominantly U.S.-based, male, and academically affiliated, with only one industry representative responding despite multiple outreach attempts. These limitations may affect the generalizability of findings to non-Western, non-English-speaking contexts, or industry contexts. Limited industry representation likely reflects disclosure constraints rather than limited engagement, underscoring the transparency gap motivating this study. Finally validity and LLM-specificity mappings reflect our structured analytic judgment rather than objective classification, and are intended as interpretive heuristics rather than sharp ontological claims.

\section{Results} \label{sec:Results}
\subsection{Descriptive statistics} \label{sec:Results_Statistics}

We provide some summary statistics about our expert interviewees, as well as the studies that interviewees had conducted and that we discussed in interviews.

\subsubsection{Sample: Experts} \label{subsec:Results_Statistics:Interviewees}

\begin{table*}[t]
    \centering
    \small
    \begin{tabularx}{\textwidth}{ l X l X }
         \toprule
         \textbf{Expert ID} & \textbf{Organization Type} & \textbf{Expert ID} & \textbf{Organization Type} \\ 
         \midrule
         Expert A & Independent Research Institution & Expert I & University \\
         Expert B & University & Expert J & University \\
         Expert C & Independent Research Institution & Expert K & University \\
         Expert D & Other & Expert L & AI Company \\
         Expert E & Independent Research Institution & Expert M & Independent Research Institution \\
         Expert F & Independent Research Institution & Expert N & Independent Research Institution \\
         Expert G & Independent Research Institution & Expert O & University \\
         Expert H & Governmental Institution & Expert P & University \\
         \bottomrule
    \end{tabularx}
    \caption{Overview of Experts}
    \label{tab:Interviewees}
\end{table*}

We interviewed 16 experts (Table~\ref{tab:Interviewees}), including 13 male and 3 female experts. Educational backgrounds included PhD ($n=7$), master's ($n=6$), or bachelor's degrees ($n=3$). Interviewed experts were spread across a range of seniority levels, with 0--5 years of work experience ($n = 3$), 6--10 years of work experience ($n = 7$), 11--15 years of work experience ($n = 3$), and 16+ years of work experience ($n = 3$). Experts were affiliated with universities ($n=6$), independent research institutions ($n=7$), government ($n=1$), an AI company ($n=1$), and other organization ($n=1$). Institutions were based in the U.S. ($n=14$), Germany ($n=1$), and unspecified ($n=1$).

\subsubsection{Sample: Human uplift studies} \label{subsec:Results_Statistics:Studies}

Experts had completed or were, at the time of interview, conducting between 1 and 6 LLM human uplift studies, with most working on just one ($n=10$). We discussed 16 distinct studies total, including those published ($n=9$), under review ($n=4$), and not-expected-to-be-published due to proprietary or security concerns ($n=3$). Studies focused primarily on biology/biological risk ($n=6$), with others spanning software engineering, cybersecurity, medicine, social sciences, and other domains. Seven studies aimed to evaluate, in some capacity, the potential for LLM systems to enable worst-case misuse risks by an attacker or threat actor. Most research teams included domain experts ($n=12$); and half included social scientists ($n=8$). All studies except one were randomized controlled trials with at least two arms (LLM access vs. control) ($n=15$). Sample sizes in discussed uplift studies ranged from under 20 to nearly 5000 participants ($\texttt{median}=110$), with only three studies exceeding 1000 participants. Studies recruited participants primarily via convenience sampling through partner organizations, social media, or targeted outreach.

\subsection{Thematic analysis} \label{sec:Results_Thematic_Analysis}

We present our thematic analysis in two stages. First, we present a structured synthesis of methodological challenges, organized across the AI evaluation lifecycle as defined in \cite{wei2025recommendationsreportingchecklistrigorous, paskov_preliminary_2025}. We annotate challenges with [C], [I], or [E] to indicate the primary forms of validity they threaten, as defined in Table~\ref{tab:validity_definitions}. We furthermore tag challenges with a star rating ($\star$ to $\star\star\star$) to indicate its degree of LLM-specificity, as defined in Table~\ref{tab:llm_specificity}. Some challenges, like intervention fidelity (C5), are distinct to LLM-based RCTs. Others are \textit{familiar in form but transformed} by properties of frontier AI systems, such as control conditions (C2) and interference (C6). Still others, such as interpretation of results across models and time (C9), are well-established in experimental design but take on distinct form in LLM uplift studies, particularly in high-stakes governance contexts. We then summarize a key subset of solutions. Bolding within quotations is added for emphasis. Table~\ref{tab:challenge_solution_map} maps challenges to corresponding solutions. 

\subsubsection{Methodological Challenges}\label{challenges}\label{results:rq}

\subsubsection*{C1 Measurement Under Constrained and Incomplete Pathway Coverage (Design) [C] $\star$}
\label{results:measurement}

 A growing body of evidence documents limitations of measurement instruments in AI evaluation \cite{raji2021aiwideworldbenchmark, hutchinson2022evaluationgapsmachinelearning, reuel2024betterbenchassessingaibenchmarks, conf/aies/RauhMMHCASMBKGB24, eriksson2025trustaibenchmarksinterdisciplinary, wallach_position_2025, noauthor_american_nodate}, with human uplift studies as no exception. Experts described challenges in defining measurement instruments that proxy real-world behavior, particularly in safety and misuse-focused uplift studies, where action spaces are wide, settings are complex and adversarial, and motivating threat models are themselves often under-specified \cite{bengio_international_2026}. Several experts emphasized concerns about the realism of experimental tasks. Expert N discussed a cybersecurity study:

\begin{displayquote}
\itshape
\textbf{N:} \enquote{One big challenge we're facing is whether or not the tasks, the actual lab environments that we're giving these proxy attackers is realistic enough of the real world. Are we representing the real cyber world in assigning these tasks? That's one big challenge. \textbf{And the implication is, if it's not realistic, then who cares about the uplift that's given?}}
\end{displayquote}

Experts noted that concerns about task realism are compounded by the need to constrain action spaces in order to make studies tractable. In domains with wide action spaces and multiple viable strategies, researchers must often focus on a subset of pathways, introducing trade-offs between granularity and coverage. Expert A remarked:

\begin{displayquote}
\itshape
\textbf{A:} \enquote{Studies need to artificially constrain [misuse scenario] pathways in order to be able to study anything at all...I think it's important that we don't lose sight of that. Say that you have specified one particular pathway and you didn't see uplift in that, \textbf{that doesn't need to necessarily mean that you're now able to rule out all the other pathways that you didn't study.}}
\end{displayquote}

Together, these challenges underscore that constrained and incomplete pathway coverage is often unavoidable in human uplift studies, but remains methodologically defensible only when task constraints are clearly specified, transparently communicated, and explicitly aligned with the research question of interest. This distinction is especially salient in governance and policy contexts, where results from a single constrained pathway may be interpreted, implicitly or explicitly, as speaking to a broader space of real-world behaviors or risks than that which the study was designed to measure.

\subsubsection*{C2 Control Conditions in AI-Integrated Environments (Design) [C] $\star\star$}\label{results:control}
Selecting appropriate control conditions, defined here as the within-study comparison group used to identify the causal effect of AI systems, emerged as a central design challenge shaping both the interpretation and comparability of human uplift studies.\footnote{Throughout the paper, we distinguish \textit{control conditions}, which are study-specific comparison groups used for within-study causal identification, from \textit{baselines}, which are reference points used to contextualize and compare uplift magnitudes across studies or over time. We further refer to \textit{human baselines} as baselines anchored to prevailing human capabilities or workflows \cite{wei2025recommendationsreportingchecklistrigorous}. We avoid using ``baseline'' to denote pre-intervention measurements, though this too is a common use case in literature.} While control specification is a general concern in randomized experiments, experts noted its unique challenge in human uplift studies, given the increasing embeddedness of AI tools in everyday workflows. In such settings, identifying a realistic and meaningful counterfactual is often nontrivial.

Experts described wide variation in control definitions in practice. Some studies restrict participants to basic internet search or non-AI tools, while others provide access to human experts or alternative software systems. Experts noted that, unlike many traditional interventions, LLMs do not replace a single prior tool but are layered onto an existing ecosystem of technologies. As such, control conditions are inherently relative to a reference point, which, if not made explicit, can lead to misinterpretation of uplift estimates. Expert P noted:

\begin{displayquote}
\itshape
\textbf{P:} In any control setting, there's going to be some technology that they have available to them, whether it be AI, or it used to be called AI but now it's no longer AI, or maybe it's some other AI tool...\textbf{it's always relative to something and maybe impacts interpretation.} Think about the coding papers, for example, that study how Copilot impacts programmers. Well, before Copilot, there was Autocomplete and TabComplete and IDEs...
\end{displayquote}

The appropriate choice of control conditions ultimately depends on the specific research question and the reference point or threshold of interest, for which experts suggest scientific consensus or stakeholder consultation may be useful. 

\subsubsection*{C3 Recruiting Populations Aligned with Specialized Research Questions (Recruitment) [E] $\star$}\label{results:recruit}
Experts described recruitment as a central constraint on external validity, particularly when research questions target specialized or high-stakes domains such as biosecurity, cybersecurity, or law. Experts described two recurring recruitment constraints. First, in many governance-, safety- and security-oriented studies, the populations of greatest interest are not directly recruitable at all. When research seeks to model malicious or highly capable threat actors, researchers must rely on proxy populations such as students, professionals, or domain experts, whose motivations, incentives, and constraints differ systematically from those of real-world adversaries. In these cases, uplift estimates depend on how well study designs, incentives, and task framing approximate relevant behaviors.

Second, in other settings, the population of interest is often difficult or costly to recruit. For example, Expert E examined AI use in legal contexts but recruited law students rather than practicing lawyers, noting that ``lawyers' time at law firms is pretty costly \ldots\ you just run up against budget constraints and have to make these trade-offs.'' Similar challenges arise in domains requiring specialized technical expertise or narrow sub-disciplinary knowledge. Researchers often trade representativeness for feasibility, sometimes favoring more curated samples to reduce variance or improve retention at the cost of smaller or less representative populations. Across scenarios, recruitment choices shape results. In policy-relevant settings, where decisions frequently hinge on rare or adversarial behaviors rather than average effects, failure to clearly communicate recruitment limitations can lead to overconfident extrapolation about populations or behaviors.

\subsubsection*{C4 Heterogeneous and Evolving AI Literacy (Recruitment) [I, E] $\star\star$}
\label{results:digital_lit}
Participant proficiency in the use of AI, or \textit{AI literacy}, emerged as a salient consideration in the design and interpretation of human uplift studies. Experts emphasized that variation in AI literacy is especially consequential, given the degree to which performance depends on users' prompts, output interpretation, and tool integration. Expert L noted:

\begin{displayquote}
\textbf{L:} \enquote{\itshape If you're selecting someone who is a complete novice, has never used GenAI, I think \textbf{you will not see much uplift just because they don't know how to use the tool}, whereas if you just give the same person six months to learn about GenAI and ask them again the same question six months down the road, they might be successful in using GenAI in a particular manner.}
\end{displayquote}

Heterogeneity in AI literacy poses distinct threats to both external and internal validity. If the population of interest is mis-specified, such that participant skill does not reflect the population of interest, uplift estimates may fail to generalize, undermining external validity. If variation in AI literacy is unevenly distributed or not controlled for within a study, it may act as a confounder, threatening internal validity by obscuring the causal relationship between AI assistance and observed outcomes.

\subsubsection*{C5 Intervention Fidelity Under Rapid Model Evolution (Execution) [C, I] $\star\star\star$}
\label{results:intervention}

Experts identified intervention fidelity as a key challenge: unlike traditional interventions, the LLM under study can change mid-experiment, silently and asymmetrically (see, e.g., \parencite[A.3][]{kapoor2025holisticagentleaderboardmissing}; \parencite{zhu2026auditingblackboxllmapis}). In human uplift studies, the intervention typically includes access to a particular AI model, often embedded within a broader tool ecosystem including system prompts, safety filters, plugins, or auxiliary tools. Experts noted that developer or provider changes to any of these underlying components, while often unflagged, can materially affect study outcomes. A lack of clear visibility into model versioning, common among experts, induces uncertainty about intervention fidelity. Expert N noted:

\begin{displayquote}
\textbf{N:} \enquote{\itshape When we started talking about this experiment and designing it, we noticed that the publicly available models were capable of running code in their environments and installing different Python tools. \textbf{For example, we are starting [now] to see a lot more refusals from the same exact model.} So the model has undergone an update and we no longer have access to that snapshot of the previous instance of that same model. And so if you run a study over a period of three months in which that model is being updated and you're unaware, \textbf{you're comparing apples and oranges.}}
\end{displayquote}

When participants are exposed to materially different model versions or configurations at the same time, this introduces unbalanced heterogeneity in the intervention, threatening internal validity by violating the assumption that treatment is consistently defined across subjects. By contrast, when the intervention changes uniformly over time, for example, due to a globally deployed model update, internal validity may be preserved, provided that participants are affected symmetrically. However, rather than estimating the effect of a single, fixed model, the study would then capture the effect of exposure to a changing system over a specified time horizon. The risk of such changes increases with study duration: brief laboratory tasks may face minimal exposure, while longitudinal or real-world studies spanning weeks or months are more vulnerable. 

\subsubsection*{C6 Interference: Spillovers and Contamination in AI-Integrated Environments (Execution) [I] $\star\star$}\label{results:interference}

Experts highlighted challenges arising from spillovers and contamination in AI-integrated environments. \textit{Spillovers} occur when exposure to AI diffuses indirectly, for example through social interaction, shared strategies, or collaboration, violating standard non-interference assumptions of RCTs. These risks are especially salient in settings with close cohort structures, such as classrooms, labs, workplaces, or training programs, where participants naturally exchange information. \textit{Contamination}, by contrast, occurs when control group participants directly access restricted AI tools or comparable systems, violating experimental protocol. As Expert A noted:

\begin{displayquote}
\textbf{A:} \enquote{\itshape I expect \textbf{cheating to be much more salient in LLM uplift studies, especially if there is an internet-only control group.} Contrast this with a clinical drug trial, where, if you're not giving the control group the drugs, they're probably not going to be able to acquire it.}
\end{displayquote}

Across both dynamics exists a trade-off: study designs that aim to improve external validity by approximating real-world, longer-term use often face heightened risks of spillovers and contamination. Short, tightly controlled studies, on the other hand, may better preserve causal identification, but fail to reflect real-world AI use.

\subsubsection*{C7 Expectancy Effects (Execution) [I, E] $\star\star$}
\label{results:expectation}

Experts discussed limits to blinding in human uplift studies: because interaction with AI systems is explicit, conversational, and central to task execution, blinding participants to AI treatment is often infeasible. As a result, observed effects may reflect not only the technical capabilities of the system, but also users’ expectations about what AI can do. In experimental psychology, such dynamics are often described \emph{expectancy effects} (e.g. \cite{doi:10.1177/1745691612463704}). Expectancy effects complicate interpretation when uplift results are used to make claims about underlying model performance or to support cross-study comparisons. Experts discussed partial mitigations -- such as blinding outcome assessors or analysts -- though such measures do not fully resolve these dynamics. Instead, the central interpretive choice lies in whether expectation-driven behavior is treated as a confound to be minimized and controlled for or as an integral feature of realistic AI use. Blinding challenges limit internal validity when uplift estimates are interpreted as isolating model capability, and external validity when results are compared across studies with different framing, populations, or deployment contexts.

\subsubsection*{C8 Documentation Under Proprietary and Security Constraints (Documentation) [E] $\star$}\label{results:documentation}

Experts described documentation --- or the reporting of evaluation results, metrics, methodological details, experimental materials, and datasets --- as a challenge to understanding evidence and advancing methodologies. While evaluations in many fields follow established documentation standards \cite{hopewell_consort_2025, noauthor_about_nodate}, human uplift studies fall short due to their relative nascency, along with proprietary and security concerns. Expert D noted:

\begin{displayquote}
\textbf{D:} \enquote{\itshape It’s often \textbf{very difficult to know what was actually done on an uplift study based on model cards.} Did they have a day to do the task? A week? Was it a multiple choice exam? An intensive planning exercise?}
\end{displayquote}

This lack of detail hampers scientific progress and appropriate interpretation of results for governance applications and beyond. While Expert B noted contractual solutions can ensure publishing rights, the feasibility of such arrangements can depend on power dynamics and dependencies between researchers, evaluators, and model developers, which shape the negotiability of terms in practice. 

\subsubsection*{C9 Interpreting Results Across Models and Time (Documentation) [E] $\star$}
\label{results:interpret}

While LLM uplift studies estimate a point-in-time effect of a specific model for a specific population on a specific task, their results often inform broader and longer-term governance questions on, for example, safeguards and deployment. The dynamics and constraints discussed in this section introduce challenges to understanding to what extent a given study’s estimates extend beyond its original conditions. Expert A noted:

\begin{displayquote}
\itshape
\textbf{A:} \enquote{There's often a conflation of what study questions are of interest. There's this overall question: how good are people at using AIs for this task right now? But often the study question that people want to answer in a safety context is: if we launch this model, and then in the next few years people become much better at using AIs, how well would they perform? ... I wish that studies could better isolate those two different questions and communicate clearly how their results should be interpreted.}
\end{displayquote}

The implications are greater when uplift results inform high-stakes decisions such as in safety cases and policy frameworks. Expert D illustrated how quickly the underlying conditions drift:

\begin{displayquote}
\textbf{D:} \enquote{\itshape Comparison over time is very difficult...if you run another study with the same group of people, \textbf{six months later, the world has probably changed in pretty meaningful ways.} People are more familiar with using LLMs...That's going to change the way that they perform.}
\end{displayquote}

Interpretive challenges further compound in cross-model comparisons. Unlike MCQA benchmarks such as MMLU \cite{hendrycks2021measuringmassivemultitasklanguage} or GPQA \cite{rein2023gpqagraduatelevelgoogleproofqa}, human uplift studies cannot be easily rerun on updated models or alternative systems without substantial recruitment and execution costs. As a result, direct comparison across model versions is often infeasible in practice. Expert O explained:

\begin{displayquote}
\textbf{O:} \enquote{\itshape If you ran a study with this model and then reviewers are like, well, why don't you try model X or Y? \textbf{It’s not a static benchmark that you can just rerun.} You’d have to recruit a new group of participants, which is not necessarily realistic.}
\end{displayquote}

Interpretive challenges similarly extend to baseline specification, where baseline refers to the reference point used to contextualize and compare uplift magnitudes across studies or over time. As decision-relevant comparison groups shift, driven by improvements in open-source models or the increasing integration of AI into everyday tools, reliance on outdated baselines risks comparisons that no longer reflect realistic conditions. Simultaneously, this dynamic risks what Expert D characterized as a ``boiling frog'' problem, in which gradual changes in reference points obscure substantial shifts in absolute capability \cite{kapoor_societal_2024}.

\subsubsection{Practical Solutions for the Field}\label{solutions}
Experts proposed a range of solutions associated with the challenges above, spanning study-level practices to ecosystem-level interventions. We map these solutions to challenges below and in Table~\ref{tab:challenge_solution_map}. In doing so, we aim to illustrate potential points of leverage rather than to advocate for any specific approach. This mapping is not exhaustive, prescriptive, or definitive.

\subsubsection*{S1 Standardized Task Libraries}
\label{solutions: task}

Experts proposed developing shared benchmark tasks for different task domains, given ``huge economies of scale'' and suggested that consortia of organizations could jointly contribute tasks for the common good without bearing full individual costs. Experts emphasized the importance of determining which tasks are representative or appropriately difficult. To this end, Expert A suggested consulting domain experts and surveying study designers, decision-makers, and relevant communities about proxy quality prior to study launch, noting that pre-registered expectations about real-world implications could help structure post-results interpretation. Expert B proposed using multiple KPIs to capture distinct dimensions of performance, while other experts similarly highlighted sub-task–level measurement as a way to preserve informative signal in complex, multi-step tasks or capabilities. More thoughtfully designed, crowd-sourced standardized task libraries can deepen and broaden measurement by encouraging systematic coverage of relevant questions and pathways and improving the validity of proxies through economies of scale.
\\\textbf{Related challenges: C1, C8, C9}

\subsubsection*{S2 Baseline and Control Selection Conventions}
\label{solutions: control_conventions}

Experts advocated for clearer conventions distinguishing baseline and control selection to improve interpretability and comparability across uplift studies. Suggested practices included explicitly characterizing the \emph{prior technology or workflow} being displaced when defining design and, where feasible, adopting standardized baseline and control bundles for common settings (e.g., academic research, professional services). Experts emphasized that both baseline and control choices should be made explicit and align with the specific reference point or decision threshold relevant to the downstream use of the results. Explicit baseline and control conventions could anchor results to a clear reference point, improving interpretability and cross-study comparability. Further, conventions could align evaluations with decision-relevant questions, support consistent documentation through shared standards, and foster crowd-sourced identification and mitigation of spillovers or contamination.
\\\textbf{Related challenges: C2, C6, C8, C9}

\subsubsection*{S3 Leveling and Accounting for AI Literacy}
\label{solutions: literacy}

Experts reported a range of strategies for AI literacy challenges, including measuring and filtering for proficiency in recruitment, stratifying randomization on prior experience, providing introductory training to participants, and controlling for AI skill level in post-hoc analysis. Accounting for AI literacy makes user heterogeneity explicit, helping align research questions with the populations and interaction modes the study seeks to capture, while potentially mitigating effects of confounding variables. Clear documentation of skill supports transparent reporting and fosters more appropriate interpretation by clarifying which user groups to which results apply.
\\\textbf{Related challenges: C4, C8, C9}

\subsubsection*{S4 Versioned Evaluation Infrastructure,  Snapshot Access, and Verification}
\label{solutions: snapshots}

Versioned AI systems and stable evaluation snapshots could allow researchers to measure, control for, or appropriately interpret intervention fidelity. Provider-side guarantees of fixed model versions for study duration, explicit version identifiers for models and system configurations, and mediated access to snapshots through secure or tiered research environments could all promote more rigorous research \cite{Casper_2024, brundage_frontier_2026}. The success of this solution hinges on coordination between developers, providers, and researchers. Such infrastructure need not imply public release of sensitive systems, but could instead support intervention fidelity, reproducibility, and interpretability under proprietary or security constraints. Emerging research on verification offers potential solutions \cite{10.1145/3716815.3729011, scaramuzza2026showcomplyshowinganything, scaramuzza2025engineeringtrustworthymachinelearningoperations, yao2026sprintrobustmodelattribution, you_verifiable_2026}.
\\\textbf{Related challenges: C5, C8, C9} 

\subsubsection*{S5 Contamination and Spillover Management}
\label{solutions: contamination}

\indent Experts proposed mitigating contamination and spillover threats through multiple approaches: monitoring LLM usage by control groups and removing protocol violators; offering post-study ``amnesty'' periods where participants can admit to violations without penalty while allowing data exclusion; physically separating treatment and control groups through staggered scheduling or dispersed recruitment; and implementing technical controls such as providing restricted Chromebooks or network-level site blocking. Several experts emphasized the importance of designing incentive structures that encourage compliance rather than relying solely on monitoring and detection systems.\footnote{Notably, in contexts where subjects lack strong performance incentives, contamination may be less problematic. This highlights a fundamental trade-off: while incentives simulate real-world motivations, they may also incentivize control groups to cheat.} Contamination and spillover management practices can mitigate interference issues and promote better documentation and interpretation of results.
\\\textbf{Related challenges: C6, C8, C9}

\subsubsection*{S6 Natural Experiments}
\label{solutions: nat_exp} 

\indent Expert B identified phased product roll-outs and staggered deployments as underutilized opportunities for generating quasi-experimental variation in human uplift studies, noting ``it's not that expensive because [companies] are going to [deploy in phases] in many cases --- you can't give [the new technology] to everybody all at once anyways. So if they coordinate with academics, I think there's a big opportunity to ... get experimental variation.'' When roll-out timing or access is determined by operational constraints such as infrastructure limits, geographic sequencing, or user tiering, rather than individual characteristics, these settings can approximate natural experiments that support causal inference under weaker assumptions than fully randomized trials. In such cases, differences in exposure may plausibly be treated as exogenous to user ability or motivation, enabling estimation of uplift effects with reduced selection bias. Natural experiments constrain the research question to real-world adoption contexts and can sidestep recruitment challenges, given built-in inclusion of decision-relevant populations. Natural experiments can also support interpretation of results across models or deployment stages.
\\\textbf{Related challenges: C3, C9}

\subsubsection*{S7 AI-Accelerated Research Methods}
\label{solutions: ai_accelerated}

\indent Experts identified AI as a promising tool for addressing scalability challenges in uplift research, highlighting its potential to reduce timeline and cost constraints. For study design, Expert L suggested using AI agents as complementary participants in pilot studies to enable rapid testing of experimental protocols \parencite[e.g.,][]{ruan2024identifyingriskslmagents, anthis2025llmsocialsimulationspromising, lu-etal-2025-toolsandbox, froger2025arescalingagentenvironments, park2026llmagentsgroundedselfreports}. Agent-based piloting could help, for example, surface brittle task designs, unintended affordances, or reward-hacking in complex multi-step settings at low cost. AI-accelerated methods may also enable rapid exploration of questions and behavior across diverse simulated human profiles under clean experimental conditions. At the same time, agent behavior may be limited in external validity and should be cautiously vetted, validated, and used as a complement to rather than substitute for human subjects research.
\\\textbf{Related challenges: C3, C4, C5, C6}

\subsubsection*{S8 Post-Hoc Analysis}
\label{solutions: analysis}
Post-hoc analytical techniques can partially mitigate internal validity challenges in human uplift studies, particularly when ideal experimental control is infeasible. Careful ex post analysis, such as deriving heterogeneous treatment effects, adjusting for non-compliance, or reweighting observations to account for imbalance or attrition, can help diagnose and, in some cases, adjust for violations of key identification assumptions. Several experts emphasized that uplift studies generate unusually rich interaction logs, which enable finer-grained post-hoc audits of participant behavior. Expert A further noted that adhering to research standards, such as explicitly estimating causal effects rather than relying solely on directional hypothesis tests, can strengthen internal validity and downstream interpretation. While post-hoc analysis cannot substitute for robust experimental design, experts viewed it as a valuable complement.
\\\textbf{Related challenges: C3, C4, C5, C6, C7, C9}

\subsubsection*{S9 Information Security Advisory Frameworks \& Tiered-Access} \label{solutions: info_sec}
\indent For studies presenting security or proprietary concerns, Expert A advocated for structured, deliberate consultation with relevant communities through advisory boards that include domain and scientific experts. For example, human uplift studies on medical tasks may draw from consensus among medical experts, while human uplift studies on bioterrorism tasks may draw from consensus among national security advisors. This approach could help balance security, proprietary, and transparency concerns by providing independent review of information hazard risks while maintaining scientific openness where appropriate. Experts framed documentation not as a binary choice between openness and secrecy, but rather as a question of granularity and audience. Tiered-access reporting provides a practical mechanism for disclosing sufficient information to support interpretation of uplift results, while responsibly managing proprietary and security risks through controlled access  \parencite[e.g.,][]{forum_fmf_2025}.
\\\textbf{Related challenges: C8, C9}

\section{Discussion: Validity Tradeoffs, Collective Action, and Governance}

Expert interviews surfaced three recurring methodological tensions in human uplift research. Each implicates construct [C], internal [I], and external validity [E], though through different mechanisms. \textit{Realism versus control}, for example, trades ecological realism against causal identification: naturalistic settings invite the spillovers (C6) that threaten causal identification, while the controlled tasks that protect it may poorly reflect real use (C1). \textit{Speed versus rigor} trades timely evidence against design depth: rigorous, externally valid studies of long-horizon tasks may require months or years from design to completion, a window during which the intervention and research question itself may change (C5, C9). \textit{Disclosure versus protection} affects the assessment of all three validity types, with proprietary and security constraints (C8) limiting the reporting and documentation needed to effectively interpret and act on evidence.

These tensions carry direct consequences for AI governance: studies designed on weak foundations or reported without sufficient context may produce false confidence or, conversely, trigger unnecessarily burdensome obligations. Well-designed and clearly documented studies, by contrast, can strengthen the evidence base for governance. Crucially, validity concerns do not operate in isolation: design choices that strengthen one form of validity often weaken another. Policy decisions should therefore rest on convergent evidence from multiple studies using complementary approaches. In governance decisions in which the relevant question shifts faster than any one study can resolve it, uplift estimates may be combined with approaches such as scenario analysis, forecasting, and methods designed to inform decisions under deep uncertainty \parencite{lempert_dmdu_2025, marchau_dmdu_2019}. In all cases, rigorous documentation to address C9, paired with tiered-access reporting (S9) where relevant, is a precondition for effective use of uplift evidence.

While some of the highlighted challenges (e.g. C1, C3, C8, C9), familiar in form, can turn to established methodological precedent for some degree of guidance \parencite[e.g.,][]{meyer_natural_1995, hopewell_consort_2025, hoffmann_tidier_2014}, a novel set of challenges stems from properties distinctive to LLM-based systems (e.g. C5 and, to varying degrees, C2, C4, C6, C7). For these challenges, existing methodological toolkits offer weaker guidance. The solutions proposed by interviewed experts represent early attempts to fill this gap. Each solution carries its own validity tradeoffs, which may be empirically explored in future work.
 
Beyond any individual uplift study lies a meta-level need for collective action. Several proposed solutions (S1, S2, S9) depend directly on field-level coordination; without this coordination, challenges C1–C9 are more likely to persist. At present, fragmented infrastructure forces research teams to repeatedly reinvent methods, while proprietary evaluations and misaligned research incentives constrain knowledge-sharing. Field-level workshops, shared infrastructure, and consensus-building processes could enable researchers to exchange lessons and co-develop novel solutions for AI evaluation and governance. Fields with deep experience in RCTs and causal inference (e.g. medicine, psychology, and economics) remain severely underleveraged within human uplift research for both familiar and novel challenges: cross-domain synergies and bridges could meaningfully advance human uplift, evaluation, and related governance efforts. In this context, public and philanthropic actors are uniquely positioned to support evaluations and initiatives that no single organization has incentives to undertake.

\section{Conclusion}

Human uplift studies, or studies that measure AI effects on human performance relative to a status quo, offer a framework for measuring the societal impacts of AI systems. By studying how AI systems impact human performance in realistic tasks, uplift studies can inform governance decisions in ways that model-centric static benchmarks alone cannot. At the same time, the application of RCT methodologies to frontier AI systems introduces distinctive challenges that complicate both study design and interpretation and, subsequently, threaten to confound downstream decision-making.

Across 16 expert interviews, we find that these challenges map onto familiar threats to construct, internal, and external validity, amplified by the properties of frontier AI systems. These trade-offs carry direct implications for policy and governance of AI systems. No single uplift study, however well designed, can provide a definitive assessment of system safety or societal impact. Policy-relevant conclusions should therefore rest on convergent evidence drawn from multiple studies with complementary designs, rather than on isolated findings.

Looking forward, advancing the rigor and usefulness of human uplift research will require more than incremental improvements in individual study design. Established disciplines hold relevant expertise and experts identified a range of novel and practical solutions, including standardized task libraries, clearer baseline and control conventions, versioned evaluation infrastructure, and structured approaches to interference, that can partially mitigate these challenges. Most importantly, however, the field requires coordination mechanisms to address collective action problems, enabling knowledge and resources to accumulate rather than remain siloed.

As human uplift studies play a growing role in informing decisions and policymaking, strengthening the methodological foundations of uplift research and clarifying what these studies can and cannot support becomes essential for responsible governance. Doing so can contribute to a coherent and policy-relevant body of evidence towards more beneficial AI futures. 

\section*{Acknowledgements}

The authors are grateful to the expert interviewees who generously volunteered their time, experience, and judgment to participate in this study. Their insights substantially informed both the framing of the research and the interpretation of its findings. The authors also thank the leadership of RAND's Center on AI, Security, and Technology (CAST) for their support of this work. RAND Global and Emerging Risks is a division of RAND that delivers rigorous and objective public policy research on the most consequential challenges to civilization and global security. This work was undertaken through CAST, which examines the opportunities and risks of rapid technological change, with a focus on artificial intelligence, security, and biotechnology. For more information, contact \href{mailto:cast@rand.org}{cast@rand.org}.

\section*{Ethical Considerations Statement}
This study was reviewed by RAND's Institutional Review Board and deemed exempt under 2024-N0632-MOD-06. The research draws on semi-structured interviews with expert practitioners and involved no experimentation, intervention, or deception. All experts provided informed consent prior to participation and were informed of the voluntary nature of the study, their right to decline to answer questions, and their ability to withdraw at any time without penalty. Interviews were conducted using secure videoconferencing infrastructure and, with expert permission, audio-recorded for transcription. Raw recordings and non-anonymized transcripts were accessible only to authorized project staff and stored on encrypted organizational systems. Recordings were deleted following transcription and anonymization. Anonymized transcripts were reviewed to remove identifying details prior to analysis or sharing with collaborators. Experts were asked how they wished to be identified in any published materials, and no quotations were attributed to individuals by name. All quotations included in the paper were cleared with experts. These procedures were designed to minimize risks of inadvertent disclosure or reputational harm while enabling transparent reporting of methodological insights.

\section*{Funding}
This research was independently initiated and conducted within the RAND's Center on AI, Security, and Technology (CAST) using income from operations and gifts and grants from philanthropic supporters. A complete list of donors and funders is available at \href{www.rand.org/CAST}{www.rand.org/CAST}. RAND clients, donors, and grantors have no influence over research findings or recommendations.

\printbibliography

\clearpage
\addtocontents{toc}{\protect\bigskip}
\addcontentsline{toc}{section}{Appendices}
\appendix
\setcounter{secnumdepth}{1}
\onecolumn

\onecolumn
\appendix
\section{Rapid Literature Review Methodology} 
\label{appendix:Query_Results}
\label{appendix:Systematic_Review}

 To identify relevant articles, we queried Google Scholar for articles dated between January 1, 2023 and June 30, 2025 that contained any one human uplift keyword and any one LLM keyword from Table~\ref{tab:Review_Keywords}.\footnote{The start date was chosen to roughly align with the public release of modern LLMs (e.g., ChatGPT's public release on November 30, 2022\cite{openai_introducing_2022}). Articles published after the end date range were not included in our rapid literature review, though we did include such articles as seeds for our interviews if they came to our attention and fulfilled relevant inclusion/exclusion criteria.} We focus specifically on studies around LLMs given the nascent state of LLM human uplift studies, their unique challenges, and the gap of research related to their methodology. As there is no standardized terminology for human uplift studies involving LLMs, the keywords are chosen to be intentionally broad. Articles containing search terms indicating that they were literature reviews rather than experimental studies were excluded (see Table~\ref{tab:Review_Keywords}). This search resulted in a total of 106 unique articles,\footnote{The search queries initially resulted in 111 results. After initial annotation by Author 5, five articles were found to be duplicates or preprints of other articles, resulting in a total of 106 unique articles.} Author 5 then conducted initial filtering/annotation per the inclusion/exclusion criteria in Table~\ref{tab:Inclusion_Exclusion_Criteria}, with review by Author 2, resulting in a final list of 10 studies meeting our criteria. Criteria ensured that studies were included in our final list of results if they 1) contained experimental results and 2) compared performance on some tasks by humans with vs. without access to LLM systems. The final 10 studies were: \cite{wilesUsingAIUpskill, nieGPTSurpriseOffering2024, choiLawyeringAgeArtificial2024, choiAIAssistanceLegal2024, roldan-monesWhenGenAIIncreases, kuchemannCanChatGPTSupport2023, schwarczAIPoweredLawyeringAI2025c, rakapChattingGPTEnhancing2024, bastaniGenerativeAIGuardrails2025, ratkovicHarnessingGPTEnhanced2025}.
 

\begin{table}[!htb]
    \centering
    \small
    \begin{tabularx}{\linewidth}{l l X }
        \toprule
        \textbf{Type} &
        \textbf{Category} &
        \textbf{Keywords}
        \\

        \midrule

        Inclusion &
        Human Uplift &
        ``human uplift'', ``randomized controlled trial''
        \\

        \midrule

        Inclusion &
        LLM &
        ``[LLM, large language model, ChatGPT, AI, artificial intelligence, AI model, artificial intelligence model, AI system, artificial intelligence system] access'', ``access to [LLM, large language model, ChatGPT, AI, artificial intelligence, AI model, artificial intelligence model, AI system, artificial intelligence system]''$^{*}$
        \\

        Exclusion &
        Literature Review &
        ``systematic review'', ``systematic literature review'', ``scoping review'', ``scoping literature review''
        \\

        \bottomrule
    \end{tabularx}

    \vspace{2pt}
    \raggedright\footnotesize{$^{*}$ The notation ``[LLM, large language model] access'' means ``LLM access,'' ``large language model access,'' etc.}

    \caption{Search terms for rapid literature review}
    \label{tab:Review_Keywords}
\end{table}

\begin{table*}[!htbp]
    \centering
    \small
    \begin{tabularx}{\linewidth}{>{\raggedright\arraybackslash}p{3cm} >{\raggedright\arraybackslash}p{3cm} X}
        \toprule

        \textbf{Inclusion Criteria} &
        \textbf{Exclusion Criteria} &
        \textbf{Details \& Rationale}
        \\

        \midrule

        Study contained experimental results &
        Study did not contain experimental results &
        We included only primary literature with quantitative/empirical results, excluding literature reviews, pre-registrations, policy documents, etc. Our intention was to examine the methodological details of original experiments, as well as to discover experts (paper authors) with expertise in conducting human uplift studies for  expert interviews.
        \\

        &
        Non-experimental studies &
        We excluded observational studies in which experimental conditions were not controlled and participant performance was not evaluated directly by researchers (e.g., studies based on survey results or observational data). Our rationale is that these studies are significantly methodologically distinct from human uplift studies and that an important aspect of human uplift studies is the study and evaluation of human-LLM interaction in task performance (see below).
        \\

        \midrule

        Experimental design attempted to compare task performance of humans with vs. without access to LLM systems &
        Experimental conditions did not compare humans with LLM system access vs. humans without LLM system access &
        We excluded human baselines (i.e., studies comparing human-only performance vs. AI-only performance) as out of scope because they are not interactive. Note, however that there are existing guidelines for conducting robust human baselines \cite{wei2025recommendationsreportingchecklistrigorous}.
        \\

        &
        The AI systems in the experiment were not LLMs &
        Although RCTs and human uplift studies have been conducted in the context of other technologies and non-LLM AI systems, we limited the scope of this paper to LLM systems. Our rationale is that LLM systems are significantly distinct from other AI systems (e.g., due to the diversity of LLM use cases, different architectures, different methods/modes of interaction, a heightened need for construct validity in LLM evaluation vs. in traditional ML contexts, etc. \cite{salaudeen_measurement_2025, feuerriegel_generative_2024, ibrahim_towards_2025, wallach_position_2025, wei2025recommendationsreportingchecklistrigorous}) and pose significantly distinct challenges in the AI context (see Section~\ref{sec:Results_Thematic_Analysis}).
        \\

        &
        Human participants were not given direct access to LLM systems &
        We excluded experiments in which participants were given access only to static LLM outputs, rather than direct access to LLM systems. Our rationale is that direct interaction with LLM systems creates important causal pathways and threat vectors that are not present when presented with static outputs only \cite{ibrahim_towards_2025}.
        \\

        \bottomrule
    \end{tabularx}
    \caption{Inclusion/Exclusion criteria for filtering of rapid literature review articles}
    \label{tab:Inclusion_Exclusion_Criteria}
\end{table*}

Our method is subject to a few limitations. Most notably, our rapid review used Google Scholar as the search engine and used relatively restrictive search terms. Google Scholar was chosen as our search tool because most recent studies on this topic have found that it has extremely high coverage for scientific articles \cite{gusenbauerGoogleScholarOvershadow2019}, with one study finding that Google Scholar had coverage of 96\% of computer science articles indexed in other databases \cite{yasinUsingGreyLiterature2020}. Google Scholar is also appropriate for our review because it indexes preprints and gray literature: nearly all of the literature related to human uplift with LLMs has been conducted within the past few years and is thus is available only as preprints and/or on arXiv. We acknowledge that Google Scholar produces issues with the search interface, lack of precision, and reproducibility \cite{halevi_suitability_2017, boeker_google_2013}.To address interface limitations (the 256-character limit on search strings), we use multiple different search queries and combine the results. 

Our LLM search terms were also limited in that they were simple string matches for specific phrases that we identified as commonly used in human uplift studies. These search terms necessarily have low recall and inadvertently exclude relevant studies (e.g., \cite{rajashekar_human-algorithmic_2024}) because even the ``human uplift'' terminology is fairly nascent in the LLM context. Our selection of these search terms was driven by the lack of standardized language describing the experimental conditions in human uplift studies; using broader search terms resulted in many thousands of results, making analysis impossible for a rapid review.\footnote{A search for ``\,`randomized controlled trial' AND `LLM'\,'' yields close to 4,000 results, while replacing ``LLM'' with ``AI'' in the query yields nearly 28,000.} Coverage in our rapid review may have thus been limited, and we interpret our results as only being broadly suggestive of broader methodological practices.

\section{Interview Methodology} 
\label{appendix:Interview_Methodology}

Expert participants were selected using a snowball sampling method \cite{parker_snowball_2019}; we defined experts as researchers who had conducted or were currently conducting human uplift studies involving LLMs that fulfilled the criteria in Table~\ref{tab:Inclusion_Exclusion_Criteria}. 
We first created a seed set of experts, beginning with the first or corresponding authors of the 10 articles from our rapid review. We added to the seed experts from major AI developers whose recent system cards indicated that they had conducted a human uplift study (OpenAI, Anthropic, Google DeepMind, and Amazon). Based on their expertise and awareness of relevant studies, Author 9 and Author 10 then provided a list of published human uplift studies involving LLMs, and the first or corresponding authors of these studies were added to the list. Author 9 and Author 10 are senior authors on this paper who have (combined) over a decade of experience conducting randomized controlled trials, including multiple human uplift studies in the LLM context.

Author 1 and Author 2 contacted the seed experts via email, inviting them to participate in a research interview and sending them an informational fact sheet about the project. Experts were asked at the end of each interview to identify one or two other individuals that they believed we should contact to participate in interviews. Author 1 and Author 2 then contacted the snowball sampled individuals and invited them to participate in a research interview. Experts were not financially compensated for participating in interviews.

Interviews were conducted on the encrypted ZoomGov platform\footnote{One expert could not access Zoom, and the interview was conducted instead on Google Meet with no recording (the interviewer took verbatim notes instead).} by Authors 1, 2, 4, and 5, all of whom have had research experience with interview methods. Each interview consisted of one interviewer and one expert interviewee. All interviews were conducted in English. We recorded audio for transcription purposes only and with the consent of the experts. 

\begin{enumerate}
    \item Information and oral consent: interviewers gave experts an opportunity to re-review the fact sheet and collected oral consent to proceed with the interview and to record for transcription purposes.
    \item Demographics: interviewers collected demographic information from the experts (summarized in Section~\nameref{subsec:Results_Statistics:Interviewees}).
    \item Study history: interviewers defined human uplift studies as ``studies [that] measure the extent to which access to and/or use of a general-purpose AI model impacts human performance on a task, relative to a baseline (think comparing human performance with human and LLM performance). Human uplift studies often employ randomized controlled trial design.'' Experts were asked how many total human uplift studies they had conducted in the LLM context, then to answer questions about the methodological design of the (most recent) human uplift study that they had conducted. This module included questions such as the number of human participants in the study, how participants were recruited, the experimental conditions, the process for choosing and validating measurement instruments, quality control and compliance enforcement measures, etc.
    \item Methodological challenges: interviewers asked the expert to identify methodological challenges encountered when designing or implementing the study. Experts were asked to focus on methodological challenges specific to the AI context. Once experts had identified a list of methodological challenges, interviewers asked follow-up questions about how each challenge could influence the validity and reliability of human uplift studies, how the experts and their research teams addressed each challenge, and what methodological options (and tradeoffs between different options) existed for addressing each challenge.
    \item Open-ended questions: at the end of the interview, interviewers asked two open-ended questions to allow experts to identify any challenges that had not yet been discussed earlier in the interview. Interviewers also asked experts to identify other individuals or organizations whom they thought we should invite to participate in the study (snowball sampling).
\end{enumerate}

The full interview script is provided in Appendix~\ref{appendix:Script}. 

Interviews were first transcribed using a privately hosted instance of the OpenAI Whisper model; these transcripts were then manually validated by Authors 1, 2, 3, 4, and 6. Our subsequent thematic analysis followed a reflexive, two-stage inductive coding approach, following the methods of \cite{zhangIEDSExploringIntelliEmbodied2025,liUnderstandingChallengesDevelopers2022, agrawalExploringDesignGovernance2021, wei_how_2024}. 

In the first stage of TA, Authors 1, 4, 5, and 6 familiarized themselves with the interviews before independently generating codes by reviewing two randomly selected interviews from our sample through a bottom-up, open coding process. Authors 1 and 2 then discussed these four sets of codes, consolidating and refining them into an initial set of codes while grouping individual codes into categories and sub-categories. Categories were informed by the stages of the AI evaluation lifecycle, defined in \cite{paskov_preliminary_2025, wei2025recommendationsreportingchecklistrigorous}. 

In stage two of TA, Authors 3, 4, 5 used Dedoose\footnote{https://www.dedoose.com} to annotate all interviews using the initial codes developed from the first stage. Each interview was independently annotated by two coders, and no coder analyzed an interview for which they were also the interviewer. Codes were iteratively adapted and refined throughout the coding process, with minor changes made as needed and discussed with Author 1 and Author 2.

\subsection*{Interview Script} \label{appendix:Script}

Before we begin, could you confirm that you received and reviewed the fact sheet sent prior to this interview? If no, would you like another copy? It's short and I can wait while you read it.

I would like to remind you that your participation is voluntary, and you should only participate if you fully understand the study requirements and risks. Do you have any questions related to participating in the study?

Do you consent to proceed?

Thanks for agreeing to be interviewed! Before we start I want to double check with you about a few things: 1) We are audio taping this interview and we will be preparing transcripts from the interviews. Is that OK with you? 2) We will be reporting themes and variation in responses across the interviews. We may include some direct quotes, but we will not include any quotes or attributions without your consent. You are free to decline to answer any question, to provide the level of detail you feel is appropriate in any response, or to respond ``I don't know'' to any question. Does that make sense to you?


\begin{table}[H]
\small
\begin{tabular}{p{0.5cm} p{0.5cm} p{\dimexpr\textwidth-1.5cm\relax}}
\toprule
\multicolumn{3}{l}{\textbf{Module: Demographics (D)}} \\
\multicolumn{3}{l}{\textbf{Estimated Time: 5m}} \\
\midrule
D & 1 & How would you like to be identified in our paper? Options are: \\
& & Full identification with name, role, and affiliation \\
& & Identification with role and affiliation but without name \\
& & Identification by only a short descriptor such as ``evaluations researcher at an AI company'' \\
& & No identification (we would identify you as ``anonymous'' or ``anonymous expert'' with expertise in human uplift studies) \\
& & \\
& & We'd also like to include demographics. These will be broad and non-identifying. You can decline to respond if you'd like. \\
\midrule
D & 2 & What is your gender? \\
\midrule
D & 3 & How many years of experience do you have? 0-5, 6-10, 11-15, 16+? \\
\midrule
D & 4 & In which country is your research organization headquartered? \\
\midrule
D & 5 & What is your highest achieved level of education? Bachelors, Masters, PhD, Other (which)? \\
\bottomrule
\end{tabular}
\end{table}

\begin{table}[H]
\small
\begin{tabular}{p{0.5cm} p{0.5cm} p{\dimexpr\textwidth-1.5cm\relax}}
\toprule
\multicolumn{3}{l}{\textbf{Module: Study Count (SC)}} \\
\multicolumn{3}{l}{\textbf{Estimated Time: 1m}} \\
\multicolumn{3}{l}{\textit{Instructions: if SC1==yes, proceed to Module 1. If SC1==no: skip to Module 2}} \\
\midrule
& & In this interview, we'll be asking you about human uplift studies in AI evaluations. Human uplift studies measure the extent to which access to and/or use of a general-purpose AI model impacts human performance on a task, relative to a baseline. Human uplift studies often employ randomised controlled trial design. \\
& & \\
& & I'm going to start by asking you a few quick questions about human uplift studies that you've been involved in. \\
& & \\
& & Have you been involved in running any human uplift studies in the past? How many? If multiple, let's begin with the most recent, substantial, and completed study. We can come back to others, time permitting. \\
& & \\
SC & 1 & Please try to keep your responses to the following questions brief, as these questions are just to give us some context on your background and what methods are currently being used in human uplift studies. \\
\bottomrule
\end{tabular}
\end{table}

\begin{table}[H]
\small
\begin{tabular}{p{0.5cm} p{0.5cm} p{\dimexpr\textwidth-1.5cm\relax}}
\toprule
\multicolumn{3}{l}{\textbf{Module: Current State (CS)}} \\
\multicolumn{3}{l}{\textbf{Estimated Time: 15m}} \\
\multicolumn{3}{l}{\textit{Instructions: conduct Module 1-3 consecutively for the first evaluation, then loop back}} \\
\multicolumn{3}{l}{\textit{through Modules 1-2 for remaining evaluations, time permitting.}} \\
\midrule
CS & 1 & Is it published? If not, do you intend to publish it? When? \\
\midrule
CS & 2 & What was the domain/subject area? \\
\midrule
CS & 3 & What was the make-up of the team that designed and ran the study (e.g., background, skillsets)? \\
\midrule
CS & 4 & What did the study attempt to measure? \\
\midrule
CS & 5 & What was the study design, and what were the control and experimental conditions? \\
\midrule
CS & 6 & What measurement instruments were used in this study, and how were evaluation data/items created or chosen? \\
\midrule
CS & 7 & How were measurement instruments tested or validated? \\
\midrule
CS & 8 & How many human participants were there in this study? \\
\midrule
CS & 9 & If there was drop-off between the beginning and end of the trial, please report both numbers. \\
\midrule
CS & 10 & How did you recruit human participants for this study? \\
\midrule
CS & 11 & What quality control measures were used to recruit participants or to ensure participant compliance in this study? \\
\midrule
CS & 12 & What statistical methods were used to analyze this study? \\
\midrule
CS & 13 & Over what period of time did the study measure outcomes? \\
\midrule
CS & 14 & Do you intend to conduct follow-up studies with this sample? \\
\bottomrule
\end{tabular}
\end{table}

\begin{table}[H]
\small
\begin{tabular}{p{0.5cm} p{0.5cm} p{\dimexpr\textwidth-1.5cm\relax}}
\toprule
\multicolumn{3}{l}{\textbf{Module: Methodological Challenges (MC)}} \\
\multicolumn{3}{l}{\textbf{Estimated Time: 25m}} \\
\multicolumn{3}{l}{\textit{Instructions: use stages for conversational guidance rather than imposed structure.}} \\
\midrule
& & What were some of the methodological challenges that you encountered when designing or implementing this study? \\
& & \\
& & For interviewees who have not conducted a study but otherwise have insight into uplift studies, what are some of the methodological challenges that researchers encounter when designing or implementing human uplift studies study? \\
& & \\
& & 1 Design: defining the evaluation's scope, purpose, structure, methodological design etc. \\
& & 2 Implementation: selecting and constructing evaluation tools, recruiting human participants, etc. \\
& & 3 Execution: data collection stage of the study, i.e., running the study itself with AI systems and with human participants \\
& & 4 Analysis: conducting statistical analysis and/or interpreting results of the data \\
& & 5 Documentation: recording and sharing evaluation results, metrics, methodological details, etc. \\
& & \\
MC & 1 & Please try to focus on challenges specific to the AI context. For concreteness, feel free to discuss challenges at any stage of the evaluation process. \\
\midrule
\multicolumn{3}{l}{\textbf{For each challenge identified above:}} \\
\multicolumn{3}{l}{\textit{Instructions: loop through MC for *each* methodological challenge identified}} \\
\midrule
MC & 2 & How does this challenge limit human uplift studies? In other words, how could it have influenced the validity or reliability of study results? \\
\midrule
MC & 3 & Is there anything about this challenge that makes it uniquely challenging in the AI context? In other words, why don't RCTs or human uplift trials in other contexts face this challenge? \\
\midrule
MC & 4 & What are ways that you or others have tried to deal with this challenge? What are the different options here, and what are the tradeoffs? \\
\bottomrule
\end{tabular}
\end{table}

\begin{table}[H]
\small
\begin{tabular}{p{0.5cm} p{0.5cm} p{\dimexpr\textwidth-1.5cm\relax}}
\toprule
\multicolumn{3}{l}{\textbf{Module: Final questions (FQ)}} \\
\multicolumn{3}{l}{\textbf{Estimated Time: 5m}} \\
\multicolumn{3}{l}{\textit{Instructions: if time remains after this module and the interviewee has conducted}} \\
\multicolumn{3}{l}{\textit{multiple AI human uplift studies, loop back to Modules 1-2 for each subsequent}} \\
\multicolumn{3}{l}{\textit{study, time permitting.}} \\
\midrule
FQ & 1 & If you could wave a wand and solve a problem more broadly in the field, what would you want solved? \\
\midrule
FQ & 2 & Is there anything else that we haven't discussed that you would like to raise? \\
\midrule
FQ & 3 & Are there any other researchers or organizations that you suggest we reach out to for this study? As a reminder, we are focused on interviewing researchers who have conducted human uplift studies/RCTs in the context of AI. \\
\bottomrule
\end{tabular}
\end{table}


Thanks for participating in this interview! Just to re-confirm after our interview, we will give you a chance to review any quotes, attributions, identification, etc. before we release this paper. We will not include any quotes, attributions, or identification without your permission. Do you have any other questions for us about this study or about your participation in this study? If you have any other questions or comments, you can contact the research team.

\FloatBarrier
\newpage
\section{Thematic Analysis Codes} \label{appendix:Codes}
\begin{table*}[!htbp]
\centering
\small
\begin{tabularx}{\textwidth}{>{\raggedright\arraybackslash}p{4cm}X}
\toprule
\textbf{Code Category} & \textbf{Description} \\
\midrule

\multicolumn{2}{l}{\textbf{Design \& Implementation}} \\
\midrule

Defining the control conditions & Defining the control condition to which experimental conditions are compared. Control conditions may include, e.g., internet search, ~2023 level AI, use of expert human hotlines, etc. \\

Defining the experimental condition(s) & Defining the treatment/experimental condition(s), including which LLM(s) to evaluate. Factors considered may include the recency of models, elicitation of models, the pace of model deployment, interaction of models in real-world, and causal mechanisms or threat models. \\

Defining the population of interest & Specifying the subset of humans to which the study results are meant to generalize. Populations could be defined by demographic, AI skill/literacy, level of motivation, expertise, intent, etc. \\

Defining the research question and causal mechanism for impact & Hypothesizing pathways by which the AI may affect human outcomes (e.g., suggestions, explanations, speed). Researchers must test mechanisms or pathways through which access to LLMs could result in risks/benefits. Even for evaluations that are measuring LLM impacts generally, the research design is often making implicit choices to prioritize measurement of particular action spaces. \\

Designing the test environment & Deciding about the test environment, including physical spaces (e.g. lab, field, etc.) and study length. \\

Developer involvement/collaboration & Interacting with developers in the design stage, including regarding contracts and model access. \\

Domain- or methodology-specific experimental design considerations & Experimental design considerations (e.g. using a within-participant, crossover, or other more complex design) to improve measurement in a specific domain or to make a certain statistical methodology possible. \\

Measurement - proxy/dependent variable choice & Designing and piloting measurement instruments such that they act as a valid proxy for the real-world outcome researchers seek to measure. \\

Measurement - task specification & Choosing the task to be measured. May include discussions of depth-breadth tradeoff, e.g. tradeoff of pre-specifying pathways of a threat model to better measure them vs. allowing for wider decision space (with poorer measurement). \\

Preventing spillovers/contamination (quality control) & Controlling for contamination, spillovers, and cheating both \textit{in} (e.g. in the lab during study hours) and \textit{out} (e.g. out of the lab during non-study hours) of the direct study environment and both within and between treatment/control groups. \\

Uplift vs. other evaluation methods & Deciding which evaluation method to use in order to measure the concept of interest, whether uplift or another method (e.g., benchmarks). Considerations may include factors such as cost, ecological validity, etc. \\

\midrule
\multicolumn{2}{l}{\textbf{Recruitment}} \\
\midrule

Participant incentives & Choosing incentive or disincentive structures to motivate participants to participate in the study, comply with study protocols, and model real-world motivations. May include financial or reputational incentives, social dynamics, research monitoring, etc. \\

Participant recruitment & Recruiting motivated participants that proxy the real-world population of interest. Considerations may include baseline AI skill level (both in terms of using LLMs and interpreting LLM outputs), domain expertise, and demographics. \\

Participant training & Training participants on relevant skillsets, including but not limited to using LLMs and interpreting LLM outputs. \\

\bottomrule
\end{tabularx}
\caption{Human Uplift Study Coding Categories and Descriptions (continued on next page)}
\label{tab:coding_categories}
\end{table*}

\clearpage

\begin{table*}[!htbp]
\centering
\small
\addtocounter{table}{-1}
\begin{tabularx}{\textwidth}{>{\raggedright\arraybackslash}p{4cm}X}
\toprule
\textbf{Code Category} & \textbf{Description} \\

\midrule
\multicolumn{2}{l}{\textbf{Execution}} \\
\midrule

Monitoring access and use of LLM & Tracking the extent to and ways in which participants use LLMs. May include tracking message/user logs and/or number of messages, tracking time spent using LLMs, recording interactions, collecting self-reported metrics, etc. \\

Monitoring spillovers/contamination & Gathering data on the occurrence and degree of spillovers, contamination, or cheating to act on and/or account for in analysis. \\

Retention of sample & Retaining a [representative] sample, especially in the case of longitudinal studies. Synonymous with preventing attrition or drop-off. \\

Use of ethical protocols & Ensuring ethical treatment and safety of participants and society, including but not limited to in security-related evaluations. Includes mitigation of info hazards, bio hazards, etc. \\

\midrule
\multicolumn{2}{l}{\textbf{Analysis}} \\
\midrule

Controls - participant characteristics & Controlling and/or accounting for participant heterogeneity -- including AI skill level -- in analysis. \\

Controls - spillovers/contamination & Controlling for contamination, spillovers, cheating etc. in analysis. \\

Grading results/analyzing data (e.g. user logs) & Effectively making use of and analyzing large amounts of data that may be unstructured or inconsistently structured (e.g. user logs) to understand causal pathways. \\

Statistical methods & Choosing and implementing statistical methods used to analyze data and results. Includes considerations around how to aggregate metrics to answer the research question (e.g. mean v top percentile). \\

\midrule
\multicolumn{2}{l}{\textbf{Documentation}} \\
\midrule

Choosing baselines and thresholds against which to compare results & Deciding on and operationalizing success/failure thresholds and comparator choices (human baseline, heuristic). \\

Interpreting results and rapid model progress & Interpreting and applying results amidst rapidly changing/deployed models and transfer/generalizability concerns. \\

Sharing of [sensitive] methodology, data, and/or results & Deciding about and acting on publication, disclosure, and reporting on methodology, data, and/or results. \\

\midrule
\multicolumn{2}{l}{\textbf{Problem/Solution Classification}} \\
\midrule

Problem v solution & Statement pertains to a problem or solution. \\
Problem & Statement pertains to a problem. \\
Solution & Statement pertains to a current or potential solution. \\
Current solutions & Statement discusses solutions that are currently -- or have been -- used. \\
Potential/ideal solutions & Statement discusses solutions that \textit{ought to be} used. \\

\midrule
\multicolumn{2}{l}{\textbf{AI Specificity}} \\
\midrule

AI Specificity: High & Statements indicating a problem is highly specific to AI. \\
AI Specificity: Low & Statements indicating a problem is general to RCTs, uplift studies, etc., in that particular problem domain -- not to the AI context specifically. \\

\midrule
\multicolumn{2}{l}{\textbf{Solution Characteristics}} \\
\midrule

Long Term Research Questions & Statements indicating a problem will require ongoing effort. \\
Perceived insolvability & Statements indicating a problem is insolvable/intractable. \\
Perceived solvability & Statements indicating problem is solvable/tractable. \\
Effective solution & Effective solution. \\
Ineffective solution & Ineffective solution. \\
High certainty of solution effectiveness & High certainty of solution effectiveness. \\
Low certainty of solution effectiveness & Low certainty of solution effectiveness. \\

\bottomrule
\end{tabularx}
\caption{Human Uplift Study Coding Categories and Descriptions (continued)}
\end{table*}

\end{document}